\documentclass{aa}  
\usepackage{graphicx}
\usepackage{lscape}
\usepackage[varg]{txfonts}
\usepackage{color}
\usepackage{natbib,twoopt}
\usepackage[colorlinks=true,
            linkcolor=blue,
            urlcolor=blue,
            citecolor=blue]{hyperref}
\bibpunct{(}{)}{;}{a}{}{,} 

\begin{document}

   \title{Model-free inverse method for transit imaging of stellar surfaces}
   \subtitle{Using transit surveys to map stellar spot coverage}

   \author{E. Aronson
          }
   \institute{Observational Astronomy, Department of Physics and Astronomy, Uppsala University, Box 516, SE-751 20 Uppsala, Sweden.\\
              \email{erik.aronson@physics.uu.se}
              }
   \date{Received tbd; accepted tbd}
%
 
  \abstract
   {We present a model-free method for mapping surface brightness variations. }
   {We aim to develop a method that is not dependent on either stellar atmosphere models or limb-darkening equation. This method is optimized for exoplanet transit surveys such that a large database of stellar spot coverage can be created.}
   {The method uses light curves from several transit events of the same system. These light curves are phase-folded and median-combined to for a high-quality light curve without temporal local brightness variations. Stellar specific intensities are extracted from this light curve using a model-free method. We search individual light curves for departures from the median-combined light curve. Such departures are interpreted as brightness variations on the stellar surface. A map of brightness variations on the stellar surface is produced by finding the brightness distribution that can produce a synthetic light curve that fits observations well. No assumptions about the size, shape, or contrast of brightness variations are made.}
   {We successfully reproduce maps of stellar disks from both synthetic data and archive observations from FORS2,  the visual and near UV FOcal Reducer and low dispersion Spectrograph for the Very Large Telescope (VLT).}
   {}

   
   \maketitle
%

\section{Introduction}
Stellar surface inhomogeneities, often called stellar spots, commonly manifest as periodic modulation of stellar light by rotation. In the early 1980s, Khokhlova, Piskunov, and Wehlau introduced a novel technique for mapping stellar surfaces and produced the first abundance maps of chemically peculiar (CP) stars (e.g., \citet{Wehlau1982SvAL}). Later, \citet{Vogt1987ApJ} coined the name Doppler imaging when they applied a somewhat simplified version of the same method to RS Canum Venaticorum (CV) stars. \citet{Glagolevskij1985PAZh} and \citet{Semel1989A&A} generalized the Doppler imaging technique to stellar magnetic fields introducing Zeeman Doppler imaging. These techniques use the fact that brightness variations (typically cool spots) on stellar surfaces produce distortions in stellar spectral lines. If a star rotates rapidly, the shape of line profiles can be dominated by rotational Doppler broadening. The line profile distortions and the positions of brightness variations on the stellar surface are correlated. A sequence of stellar spectra during rotation can be used to reconstruct a map of brightness variations on the stellar disk. In Zeeman Doppler imaging, polarization measurements are added to this to simultaneously measure magnetic field variations, which are typically spatially connected to brightness variations \citep{Donati1997MNRAS}.  

These techniques have frequently been successfully used to produce detailed maps of stellar surface brightness and magnetic field variations (\citet{Kochukhov2004A&A, Barnes2015ApJ, rosen2015ApJ, Kunstler2015A&A, Silvester2015MNRAS} and many others). The spatial resolution of Doppler imaging is fundamentally limited by the ratio of projected rotational velocity of a star to the spectral resolution of observations \citep{Piskunov1993PASP}. This restricts the applicability of Doppler imaging to moderate and fast-rotating stars with $V sin (i) > 15$ km/s. The estimate for spatial resolution works under the assumption of a dense and homogeneous phase coverage. In practice, this assumption requires long and tedious observing runs that make Doppler imaging impractical for extensive surveys. We present an alternative technique for producing a database of stellar spot coverage here. While our method cannot produce maps of the entire stellar disk, it has the advantage of not requiring any dedicated observing time. Instead, observational data from existing exoplanet transit surveys can be used to create a large database of stellar spot coverage.

Stellar spots have frequently been successfully discovered in light curves from exoplanet transit events (\citet{Pont2007A&A, Sing2011MNRAS, Carter2011ApJ, Sanchis-Ojeda2012Natur, Valio2017ApJ} and many others). We here present a new method for detecting and mapping these spots. The method, described in detail in Sec.~\ref{method}, is model-free and optimized for survey-type transit observations where stars are observed during multiple consecutive transit events at a stable signal-to-noise ratio (S/N), and where orbital parameters and size ratio of planet to star are known with good accuracy from previous studies. 

The method is tested in Sec.~\ref{resutls}, first on synthetic data similar to what we expect from future exoplanet transit surveys, and then on ESO archive observations from a medium-resolution spectrograph. While these data are not from a survey, they provide transit light curves of sufficient quality to map stellar disks, thus showing that the method is capable of recovering the brightness variations without the need for specialized observations or any prior knowledge of the system other than orbital parameters and planet-to-star size ratio. Finally, we compare this method to a different method for recovering stellar spots in transit light curves: the PRISM code \citep{Tregloan-Reed2013MNRAS}. We do this by applying our method to identical data as were previously used to test PRISM.
        
\section{Method} \label{method}
\subsection{Overview}
This transit imaging technique attempts to reconstruct maps of stellar surface brightness variations by using the shape of transit light curves. We do this in three steps. First we establish the radial brightness variations of the star, excluding potential surface inhomogeneities such as spots (i.e., the stellar specific intensities). This is then used to create a map of the brightness distribution on the stellar surface of a spot-free star. Based on this map, we can simulate transits and create synthetic light curves. We search for discrepancies between simulated light curves and the observed light curve. If a planet transits a region of the star that is brighter or darker that what was expected from the spot-free case, the shape of the light curve is affected. For example, if the planet transits a dark spot, this can be seen as a small flux increase in the light curve. If the observed light curve has such features, we adjust our map of the star accordingly. The aim is to find a brightness map of the star that can produce synthetic light curves that match the observations. When this is achieved, we filter the map through a regularization scheme that suppresses high-frequency oscillation in the final solution. This allows us to avoid fitting observational noise. This method is limited to mapping only the region of the star that the exoplanet crosses, that is, the transit chord. Overall, this whole process can be completed in approximately 30 minutes of computing time for a single typical transit. Because each observed transit event is treated separately during the mapping procedure, this can easily be parallelized, therefore the method is well suited for analyzing large data sets such as thousands of light curves from large transit surveys.   

\subsection{Spot-free specific intensities} \label{SpotFreeSpecificIntensity}
The first step in the process is finding the radial brightness variation that we could expect from a spot-free star. We propose using a sequence of transit light curves from the same system. First, the light curves were phase-folded using orbital data from previous studies of the same system. Phase-folding assumes that the stellar rotation will bring different parts of stellar surface below the planet shadow in the same phase of the transit. This is a reasonable assumption: for example, the center of the solar disk will shift by the diameter of Jupiter in about 24 hours. The phase-folded light curve will consist of points scattered around a median curve. This curve can be recovered using various techniques, such as median filtering. The process can be helped by imposing additional constraints such as monotonicity and symmetry relative to the transit center. This step will create a single light curve with a high S/N, where temporal brightness variations in the light curve are removed. 

\begin{figure}
        \centering
        \includegraphics[width=\hsize]{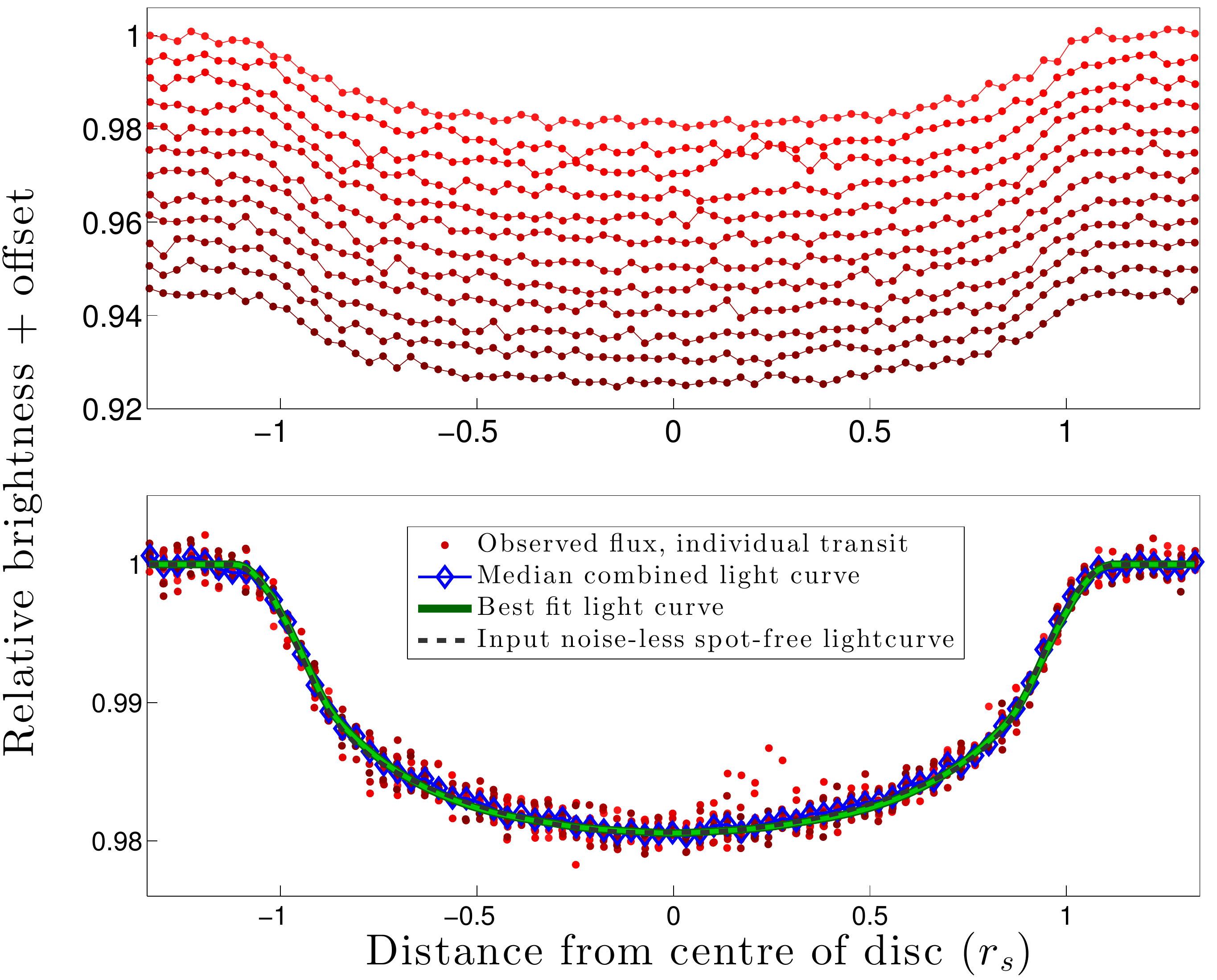}
        \caption{Sketching the method for mapping stellar surfaces, part 1: Spot-free light curve.\newline
                Upper panel: 12 consecutive (synthetic) transit light curves. Noise levels are comparable with what is expected to be routinely achieved by the Transiting Exoplanet Survey Satellite (TESS), see Sec.~\ref{Test of method on simulated data}.\newline
                Lower panel: 12 transits phase-folded (red dots) and median-combined (blue diamonds). The best fit to the median-combined light curve is shown as a solid green line. The input model used to synthesize the light curves is shown as a dashed black line.}
        \label{quietStarCreation}
\end{figure}

Using this median light curve, the specific intensities for the unspotted star can be recovered. We used the method of \citet{Aronson2018AJ}, where we demonstrated that intensity dependence on the limb distance can be uniquely defined by making two simple assumptions that are always true for a main-sequence star. These are that (1) intensity decreases toward the limb, and (2) the rate of decrease increases toward the limb. We also expect the stellar disk to be centrally symmetric, and thus the first moment of intensity must be equal to the stellar flux. No additional model or prescription of the limb darkening is required. We call the recovered intensities $I_{ref}$, interpreting them as matching a stellar surface without any spots. Fig.~\ref{quietStarCreation} shows a typical data set and the reference transit curve derived from the reconstructed specific intensities.

\subsection{Mapping the stellar surface} \label{MappingStellarSurface}
\begin{figure}[h]
        \centering
        \includegraphics[width=\hsize]{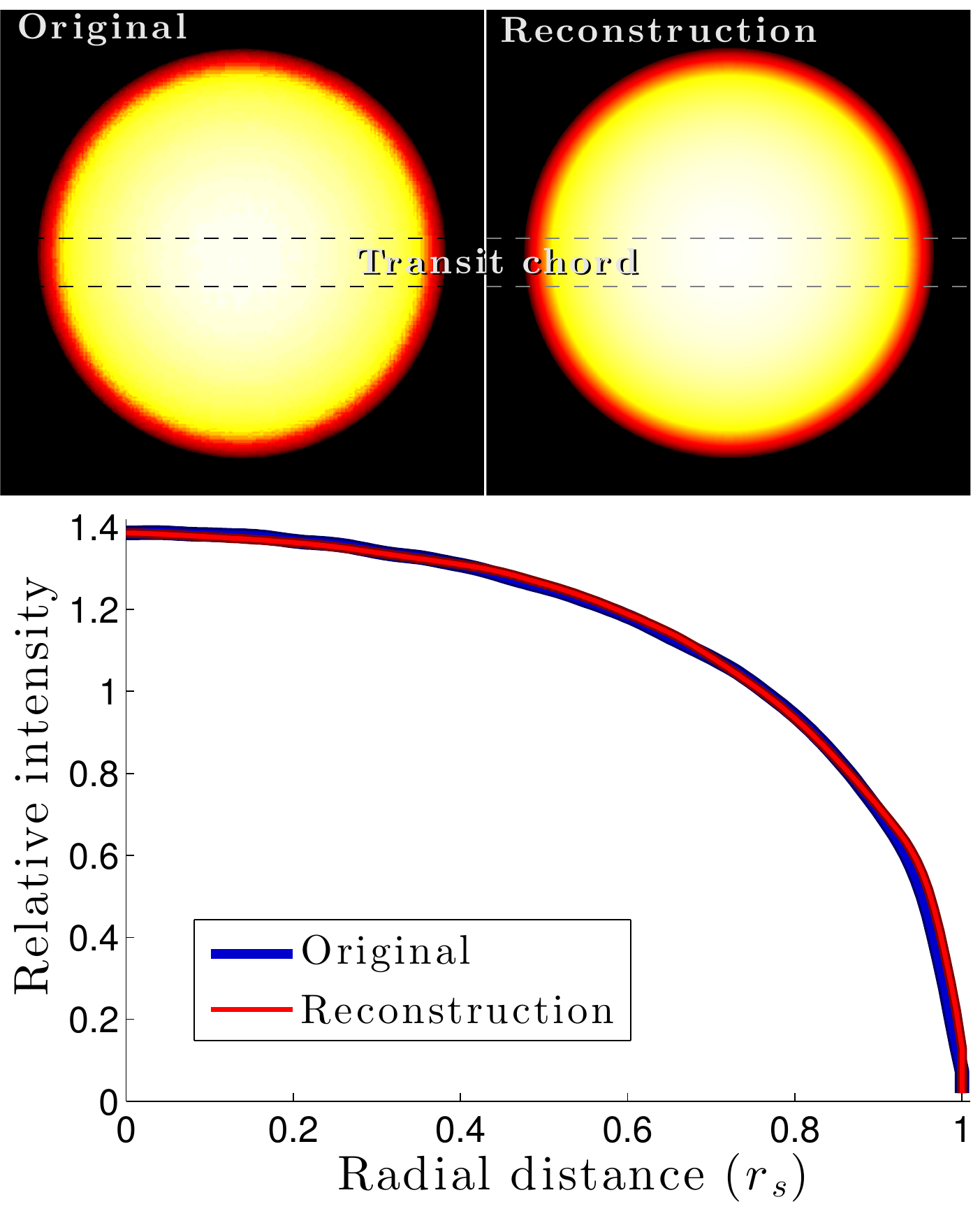}
        \caption{Sketching the method for mapping stellar surfaces, part 2: Creating the initial guess for stellar surface brightness distribution.\newline
                Upper row: Maps of the spot-free reference stellar surface. The left figure is the original that was used to create the synthetic observations that $I_{ref}$ was then fitted to. The right figure is the reconstruction of the stellar surface using $I_{ref}$. \newline
                Lower panel: Stellar specific intensities. The best-fitting specific intensity to the median-combined light curve ($I_{ref}$) is shown as a red line. This function was used to reconstitute the stellar map (upper right panel). The "true" original specific intensity of the star is shown as a blue line.}
                \label{starImageCreate}
\end{figure}

        We fit a stellar brightness map to each individual light curve from our sequence of transits. To do this, we required a first guess for the stellar surface. Using the spot-free reference intensities ($I_{ref}$), we created a two-dimensional radially symmetric image of the star. This image was divided into two parts: the region that is occulted by the exoplanet at some point during the transit, and the region that is never occulted by the exoplanet, see Fig.~\ref{starImageCreate}.
        
Using this initial guess for the stellar disk, we then synthesized the light curves, which we then compared to the observed light curve and searched for discrepancies between them. Light curves were synthesized by summing the contribution from all pixels on our stellar disk. As the planet moves across the star, it blocks some stellar light. This was simulated by creating a round elongated mask, $P$, that corresponds to the flux blocked by the planet during an exposure. The size of the planet and its position on the stellar disk at a given exposure were assumed to be known with sufficient precision. The mask was elongated in order to describe the motion of the planet during each exposure. It is the sum of many instantaneous round masks shifted along the trajectory of the planet. The flux was set to 0 where stellar light was always blocked by the planet during an exposure, and to 1 where light is never blocked. This allowed us to block the flux contribution from the occulted region of the star. In Eq.~\ref{createExp} we show how this is done mathematically. Here $f_s$ is the synthetic flux, $x$ and $y$ mark the position of a pixel and $\phi$ the phase of the planet center during the exposure. The mask $P$ is placed at a position corresponding to the location of the planet during the actual observations, which we obtained from the planetary orbital parameters and timing of each exposure. $f_s$ was computed for all observed phases. This provided a synthetic transit light curve that we normalized such that the out-of-transit flux equals one, as shown in Eq.~\ref{createExpNorm}. In this formula, $F_s$ is the normalized flux and $\phi_0$ can be any point outside the transit,

\begin{align} 
        & f_s(\phi) = \sum_{x, y} \Big( S_u(x, y) + S_o (x,y) \cdot P(x,y,\phi) \Big) \label{createExp} \\
        & F_s(\phi) = \frac{f_s(\phi)}{f_s(\phi_0).} \label{createExpNorm}
\end{align} 

\begin{figure}[h]
        \centering
        \includegraphics[width=\hsize]{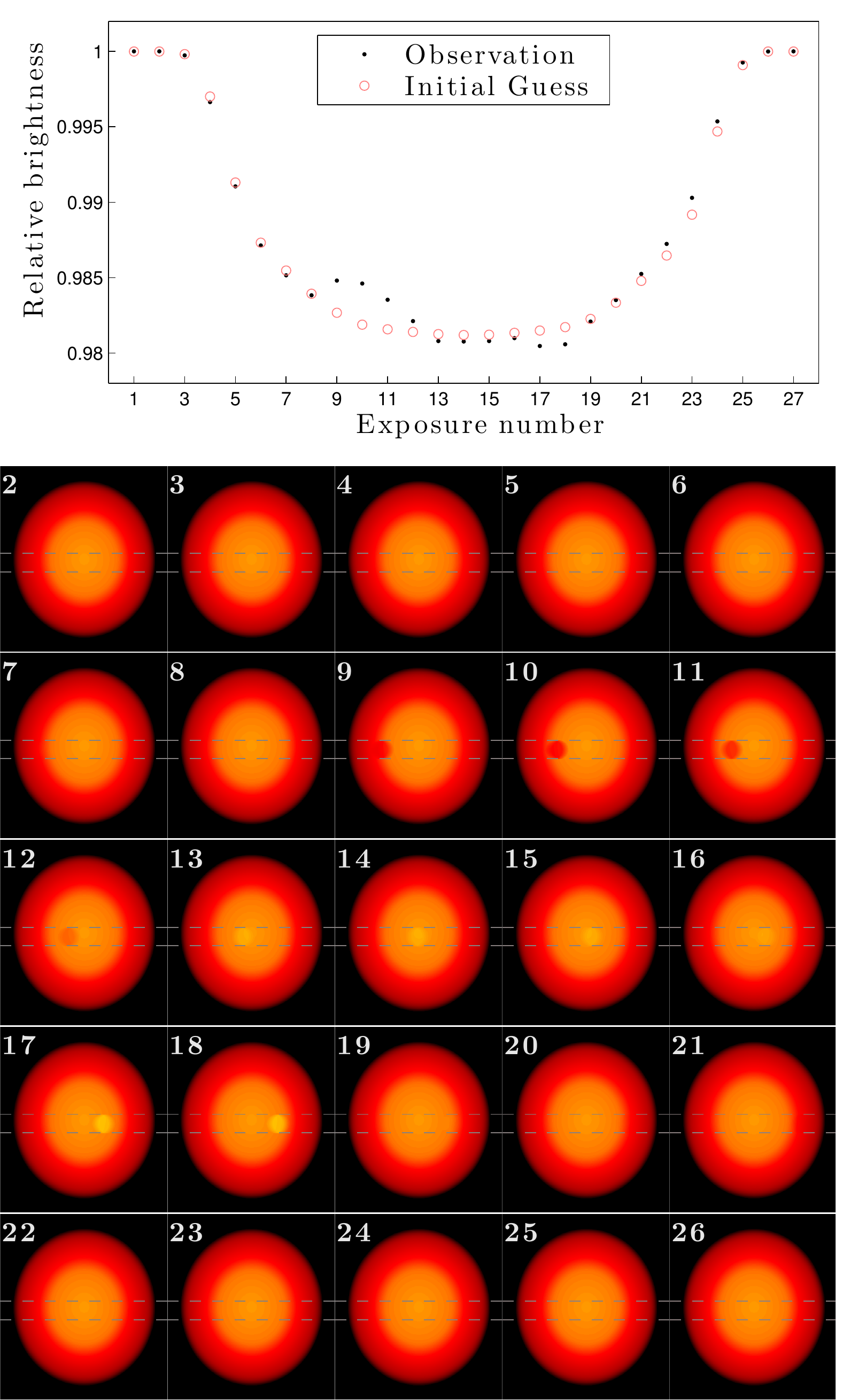}
        \caption{Sketching the method for mapping stellar surfaces, part 3: Mapping individual exposures. \newline
                Upper panel: Light curves that this sketch is based on. The observed data are shown as black dots (there is no noise in this example). The synthetic light curve produced by the initial guess for the stellar surface is shown as pink circles.\newline
                Lower 5 rows: 25 maps of the stellar surface. Each shows the updated guess for the brightness distribution based on only the one exposure that corresponds to the number in the upper left corner of each figure. Dashed gray lines show the transit chord. \newline
                For example, the predicted brightness (pink circles) in exposure 11 is lower than the observed brightness (black dots). This is interpreted as a dark spot on the stellar surface below the planet. In the updated map corresponding to exposure 11, the brightness of the stellar surface at the position of the planet in this exposure is decreased.}
        \label{starExampleMethod1}
\end{figure}

\begin{figure}[h]
        \centering
        \includegraphics[width=\hsize]{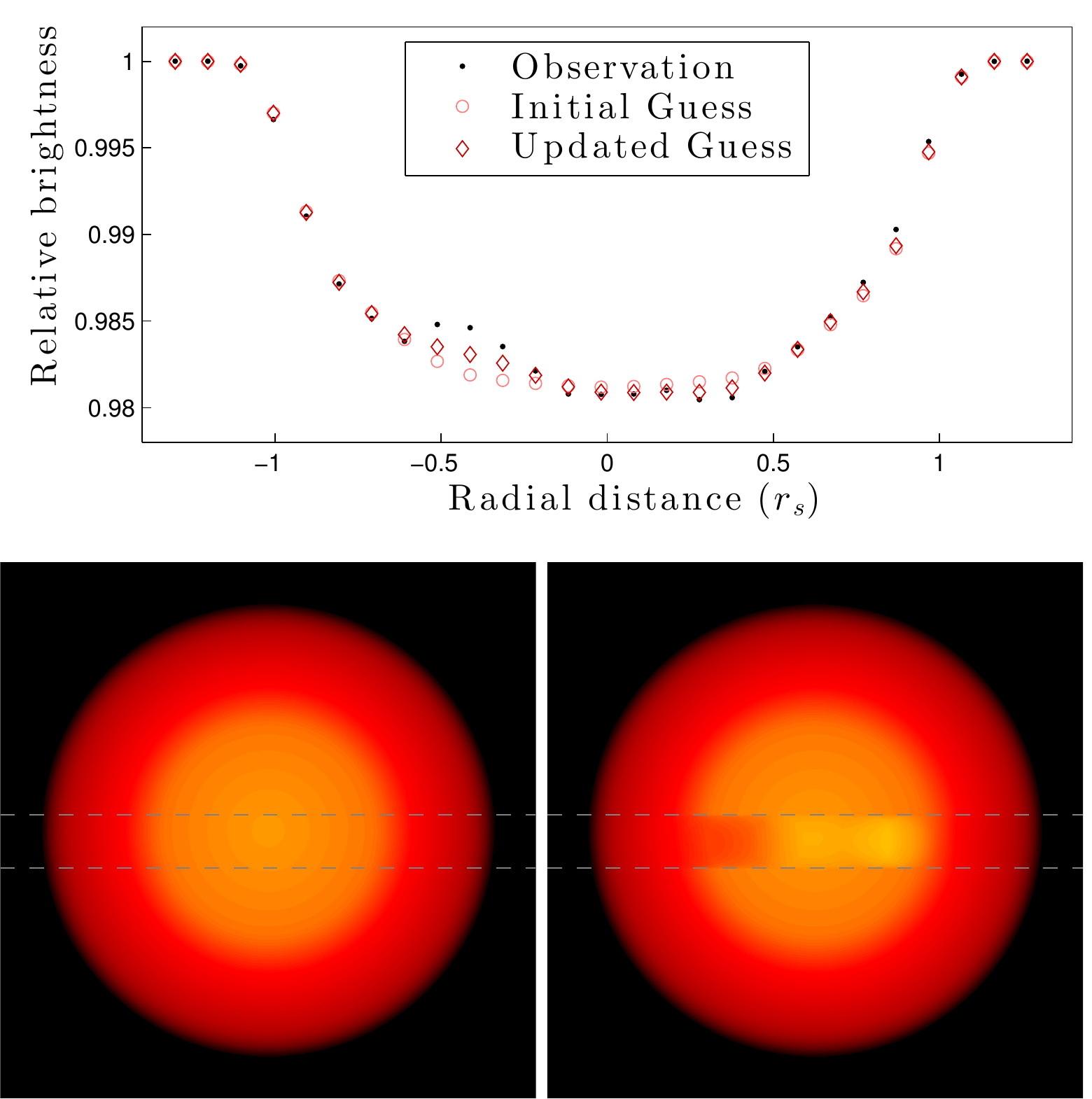}
        \caption{Sketching the method for mapping stellar surfaces, part 4: Averaging stellar maps. \newline
                        Upper panel: The observed light curve is shown as black dots (this is the exact same data as are shown in Fig.~\ref{starExampleMethod1}). The light curve that the initial guess for stellar surface produces is shown as pink circles. The light curve that the updated map produces is shown as red diamonds.\newline
                        Lower row: Maps of the stellar surface. \newline
                        To the left is the initial guess for the stellar surface, based on $I_{ref}$. Dashed gray lines shows the transit chord. \newline
                        To the right is the updated guess for the stellar surface created from averaging the maps from each individual exposure from Fig.~\ref{starExampleMethod1}. The contrast of the spots has been increased for better visibility.}
        \label{starExampleMethod2}
\end{figure}

Now the differences between the true observed light curve ($F_t$) and the synthetic light curve ($F_s$) were calculated. The discrepancies between them were used to update the map of star. This was done in two steps: First we considered the occulted part of the star. We searched for local deviations from the observed light curve. When the synthetic light curve had a lower flux than the observed light curve, this was interpreted as the planet occulting a darker part of the stellar disk than what was expected from the initial guess of the stellar surface. When the synthetic light curve had a higher flux than the observed light curve, the part of the star occulted by the planet should be comparatively brighter instead. We updated the guess for the stellar surface by examining each exposure individually. We increased or decreased all pixels of the stellar surface that were occulted by the planet during this exposure. The amount of increase or decrease was proportional to the difference between synthetic and observed light curve at this exposure. This process is sketched in Fig.~\ref{starExampleMethod1}. Each individual exposure generates a unique map of the star. In this example, the observed data show signs of stellar spots, but the initial guess has no spots. The effect of this discrepancy is strongest in exposures where the planet transits the spots, and therefore the corresponding surface map updates include dark or bright regions at the positions of the planet as it crosses the spots. After we updated the maps for each exposure, they were combined into a single updated map that takes data from all exposures into account. For this we averaged over all maps, see Fig.~\ref{starExampleMethod2}.

This process can be expressed mathematically as in Eq.~\ref{updateGuess}. Here the previous guess for the stellar surface is $S_{o_{n}}$ and the updated guess for each exposure is $s_{o_{n+1}}(\phi)$. We note the difference in terminology between uppercase $S,$ which represents the stellar surface taking all exposures into account (Fig.~\ref{starExampleMethod2}), and lowercase $s,$ which represents the surface map corresponding to a single exposure (Fig.~\ref{starExampleMethod1}). In this equation, $c$ is a free parameter that controls the amount of brightness increase or decrease. Ideally, the brightness shift should only be a fraction of what is needed for the synthetic light curve to match the observed light curve. The reasons for this are twofold. First, the part of the star that is occulted by the planet in one exposure to the next will have some overlap. Adjusting the brightness of the occulted part in one exposure may affect the flux of neighboring exposures. We therefore did not adjust the stellar maps to perfectly match the observed value in a single step. Doing so would induce errors in neighboring exposures. Instead, we converged on the observed flux through an iterative approach as detailed below.

\begin{align}  \label{updateGuess}      
        s_{o_{n+1}} (x,y,\phi) = & S_{o_{n}} (x,y) \cdot \\ \nonumber
        & \Big( 1 - \big( 1-P(x,y,\phi) \big) \cdot \big( F_t(\phi) - F_s(\phi) \big) \cdot c \Big)     
\end{align} 

\begin{figure}[h]
        \centering
        \includegraphics[width=\hsize]{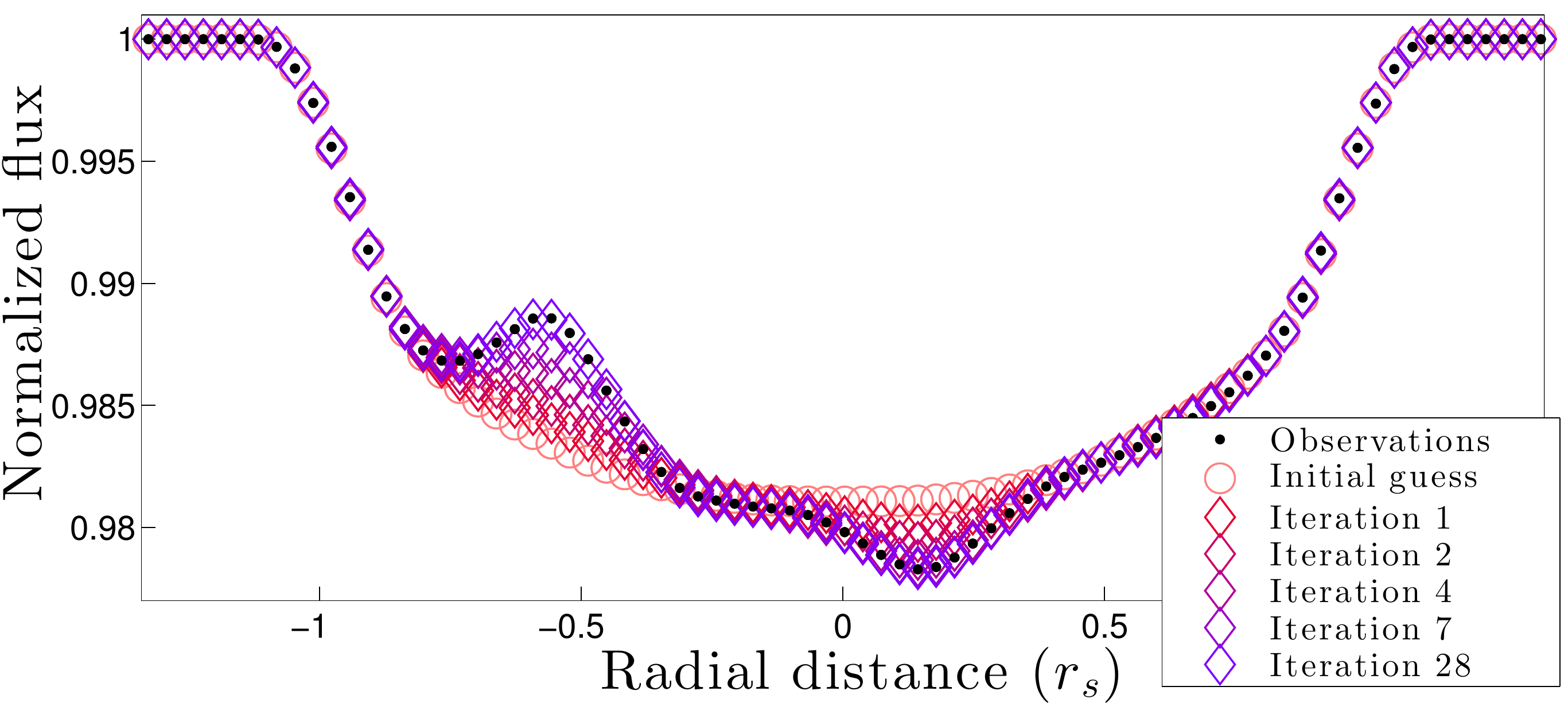}
        \includegraphics[width=\hsize]{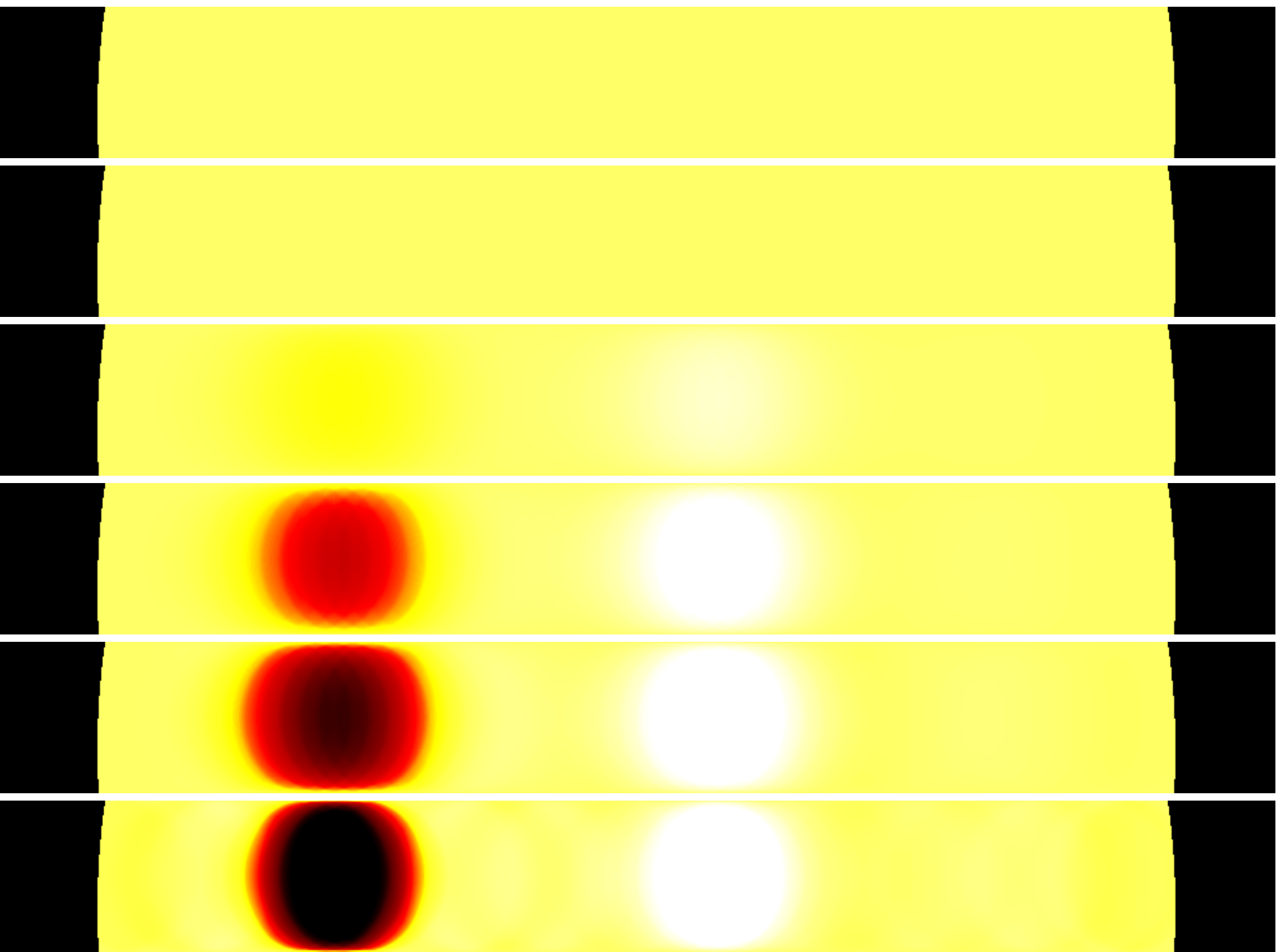}
        \caption{Sketching the method for mapping stellar surfaces, part 5: Converging on the solution.\newline
        Upper Panel: Light curves. The observed light curve is shown as black dots. The initial guess based on the spot-free map of the stellar surface is shown as pink circles. Diamonds from red to purple show the light curves at five stages in the iterative process, which are chosen to show the progress as clearly as possible. We start just after the first iteration (red) and end when convergence is found (purple). \newline
        Lower panels: Maps of the stellar surface that were used to synthesize the light curves are shown in the upper panel. Only part of the stellar disk is shown, the transit chord. From top to bottom: Initial guess (i.e., spot-free reference case), and iterations 1, 2, 4, 7, and 28.}
        \label{showProgress}
\end{figure}

The second reason for choosing $c$ such that it prevents the synthetic light curve from matching the observed curve in a single iteration is that part of the discrepancies between observed and synthetic light curves may be due to discrepancies in the overall brightness of the star, including the unocculted part. We explored this by creating a synthetic light curve using the updated map of the stellar disk. If this light curve had a different overall transit depth than what was observed, the brightness of the entire stellar disk, both the occulted and the unocculted parts, were increased or decreased by the required amount to match the observed transit depth as close as possible by minimizing the square differences between them. Now the next iteration of stellar surface mapping can be started using the updated map as the starting point. We repeated the steps detailed in this section until updating the map of the star no longer improved the fit to observations. Eventually, we converged on a map that produced a synthetic light curve that matched observations almost perfectly. This iterative process is sketched in Fig.~\ref{showProgress}.

\subsection{Regularization} \label{RegularizationMethod}
The process presented above in Sec.~\ref{MappingStellarSurface} is capable of reproducing nearly any observed light curve. We note, however, that physically, we are unable to resolve details smaller than the planet size plus planet motion during each exposure. Higher frequency temporal variations of brightness are presumably due to observational noise. Thus, we introduced a smoothing step based on a first-order Tikhonov regularization \citep{TikhonovArsenin77} to avoid unrealistic variations of specific intensity. This regularization suppresses oscillations around a mean value by finding a compromise between the fit to the data and the smoothness of the solution. Spatial resolution is traded for reduction of extensive noise-induced oscillations. We argue that this is a reasonable approach because sharp edges in the brightness variations on a stellar surface are not expected.

\begin{figure} 
        \centering
        \includegraphics[width=\hsize]{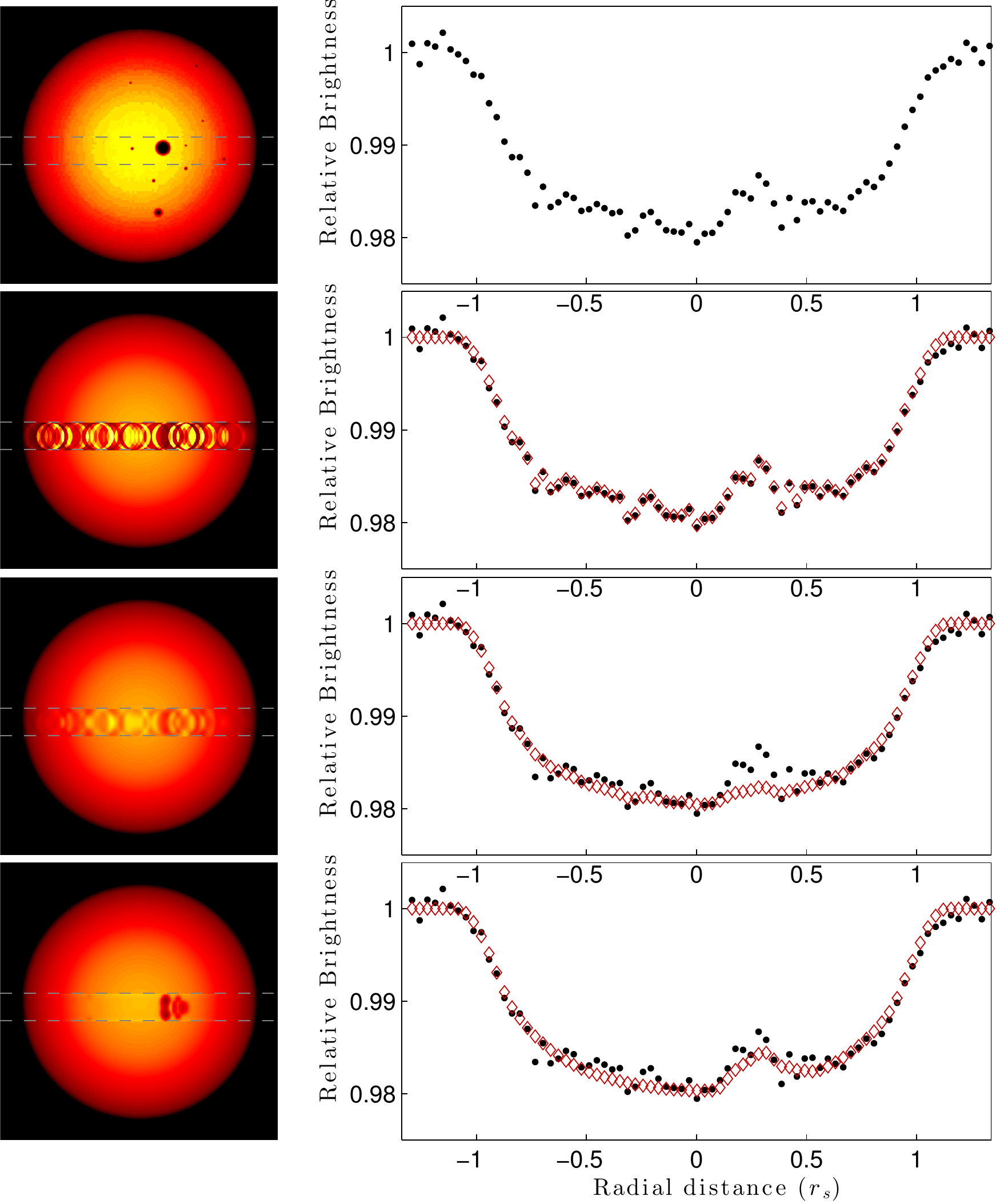}
        \caption{Sketching the method for mapping stellar surfaces, part 6: Regularization.\newline
                Rows from top to bottom are divided into a left and a right part. The left part shows a map of the stellar surface in which dashed gray lines outline the transit chord. The right part shows the corresponding light curves. Black dots show the observations, and red diamonds show the synthetic light curve that the surface map to the left produces. \newline
                First row from the top: To the left is the stellar map that was used to simulate observations. Our aim is to recover a map that resembles this. To the right is the observed light curve, including observational noise. \newline
                Second row: Recovered stellar map and corresponding light curve without any regularization. This is the end product of the process described in Sec.~\ref{MappingStellarSurface}. In this case, the noise has been interpreted as bright and dark regions on the stellar surface. \newline
                Third row: Recovered stellar surface and corresponding light curve after applying regularization with a flat weight-map. Variations due to noise have been reduced by the regularization, but can still be seen as faint bright and dark regions. The true spot is faint because the regularization causes the solution to become smoother. \newline
                Forth row: Recovered map and corresponding light curve after regularization with a dynamic weight-map, where weights are chosen such that they are low in regions where deviations from the reference case are within what can be expected from the noise levels. Now the bright and dark regions due to noise are no longer visible, and the dark region corresponding to the true spot is clearly visible, although its shape is slightly altered.}
        \label{showRegul}
\end{figure}

We introduced this regularization through Eq.~\ref{Tikho1}. Here, $S_{reg}$, $S_{o}$ , and $S_{ref}$ are brightness maps of the transit chord. $S_{o}$ is the original recovered version, that is, the end product of the iterative fitting procedure described in Sec.~\ref{MappingStellarSurface}. $S_{reg}$ is the regularized version of the same, which is created during this regularization step. $S_{ref}$ is the spot-free surface map produced from the reference specific intensities. $x$ and $y$ are pixel coordinates. $W$ are weights, and $\Lambda$ is a free parameter that controls the amount of regularization. The first term in the equation strives to keep the function ($S_{reg}$) close to the original (unregularized) stellar surface map ($S_{o}$). The second and third terms forces $S_{reg}$ to be smooth by forcing the differences between neighboring pixels to be small (the second term controls derivatives in the x-direction and the third term in the y-direction). The solution to the equation is found by searching for the function $S_{reg}$ that minimizes $\Phi$. We find this by setting derivatives of $\Phi$ to zero and linearizing the second and third terms. This creates a linear system of equations that is straightforward to solve. We demonstrate how regularization is used and how it can influence the recovered map in Fig.~\ref{showRegul}.
        
The brightness map is not a flat function because the stellar surface is curved. Regularizing the surface maps directly would therefore introduce artificial flatness in the overall brightness distribution of the star. We avoided this by applying the regularization filter on the differences between the recovered map to the reference map; that is, $(S_{reg} - S_{ref})$ and $(S_{o} - S_{ref})$. Thus the overall curvature of the stellar surface was persevered, 
        
\begin{align} \label{Tikho1}
        \Phi = \sum_{x,y} \Bigg[ & W(x,y) \cdot \Big( \big(S_{reg}(x,y) - S_{ref}(x,y) \big) - \\ \nonumber
        & \big(S_{o}(x,y) - S_{ref}(x,y) \big) \Big) \Bigg]^2 + \\ \nonumber
        \Lambda \sum_{x,y} \Bigg[ & \frac{d \big(S_{reg}(x,y) - S_{ref}(x,y)\big)}{dx} \Bigg] ^2 + \\ \nonumber
        \Lambda \sum_{x,y} \Bigg[ & \frac{d \big(S_{reg}(x,y) - S_{ref}(x,y)\big)}{dy} \Bigg] ^2 = min
\end{align}

The relative importance of the first term versus the second and third terms corresponds with fit-to-data versus smoothness-of-solution. This trade-off is controlled by the regularization parameter $\Lambda$, which can be tuned to match the overall data quality. With higher values of $\Lambda$, the second and third terms (which strive to make the solution smooth) become more important for minimization of $\Phi$. If noise levels are high, a large regularization parameter can suppress extensive oscillations, which means that as $\Lambda$ increases, the likelihood of false detections decreases. At the same time, however, the sensitivity of the method also decreases with $\Lambda$. By using simulated data with known S/N and regularization parameter, the expected number of false detections, the threshold for the spot faintness that can be detected at a given regularization parameter, and the data quality can be estimated. 

We used the following procedure to determine the optimum value of the regularization parameter. First, we defined the upper boundary for the number of false spots recovered in the analysis, for example 5\%. We ran multiple realizations of synthetic observations withou spots and a given S/N. This was fed through the transit mapping scheme, and we counted the number of maps recovered with spots for a given value of the regularization parameter. We searched for a minimum value of the regularization parameter that allowed only 5\% or less of the simulations to show detectable spots. For this purpose, we defined spots to be present in a map that differs from the reference case by more than 1\%. This allowed us to find the best-fit regularization for a given S/N and requested upper boundary for a false detection. This best-fit regularization parameter is unique for each S/N.

The weights ($W(x,y)$) have an individual value in each pixel, and they were used to increase or decrease the importance of regularization in local regions on the reconstructed map. When the synthetic light curve deviated from observation only within what could be expected from a known S/N, the weights in this regions were decreased so that regularization dominated. Thus we avoided fitting noise and mapped only deviations from the spot-free case that were stronger than the noise levels. Weights can be created by starting with a flat weight-map and updating it after producing a first unregularized map. We decreased weights in regions of the stellar map that contributed to the light curve where the deviations from the spot-free case were within noise levels. In regions that contributed to the light curve where the deviations were outside what could be expected from the noise level, weights were instead increased. This produced a weight map that suppressed small deviations that are likely due to random noise and simultaneously avoided suppressing significant deviations that are likely due to brightness variation on the stellar surface. 

\section{Results} \label{resutls}
In this section we present some tests of our method. The first test is to determine the reconstruction quality of the brightness map using synthetic observations (Sec.~\ref{Test of method on simulated data}). The second test is based on data obtained with the ESO Very Large Telescope (VLT), which allows us to compare the proposed procedure with all the complications of actual observations (Sec.~\ref{Test of method on individual transit}). The third test is a comparison with a different method for reconstructing stellar spots using transit light curves (Sec.~\ref{prism_comp}).

\subsection{Test of the method on simulated data} \label{Test of method on simulated data}
\begin{figure*}[h]
        \centering
        \includegraphics[width=\hsize]{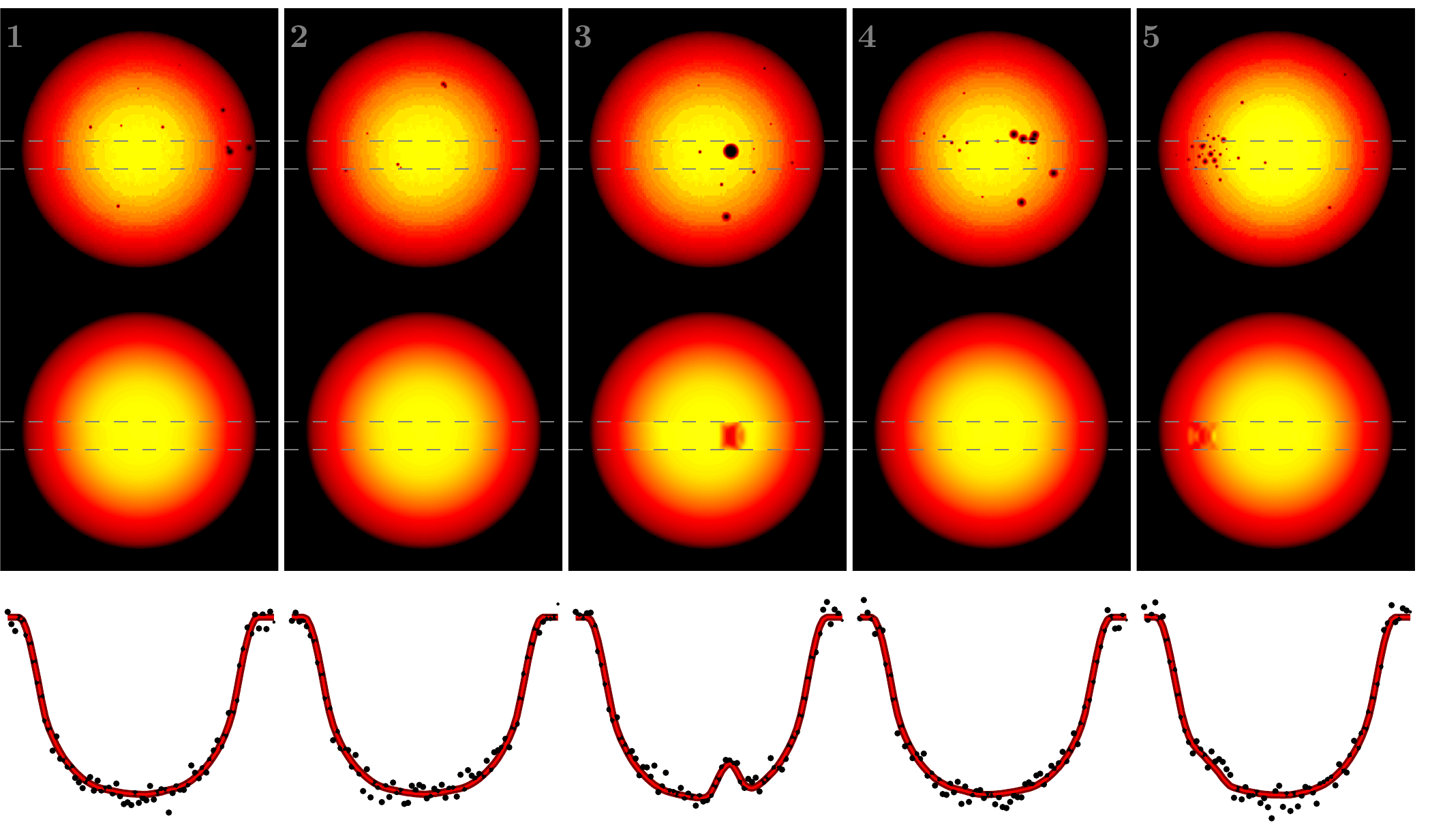}
        \includegraphics[width=\hsize]{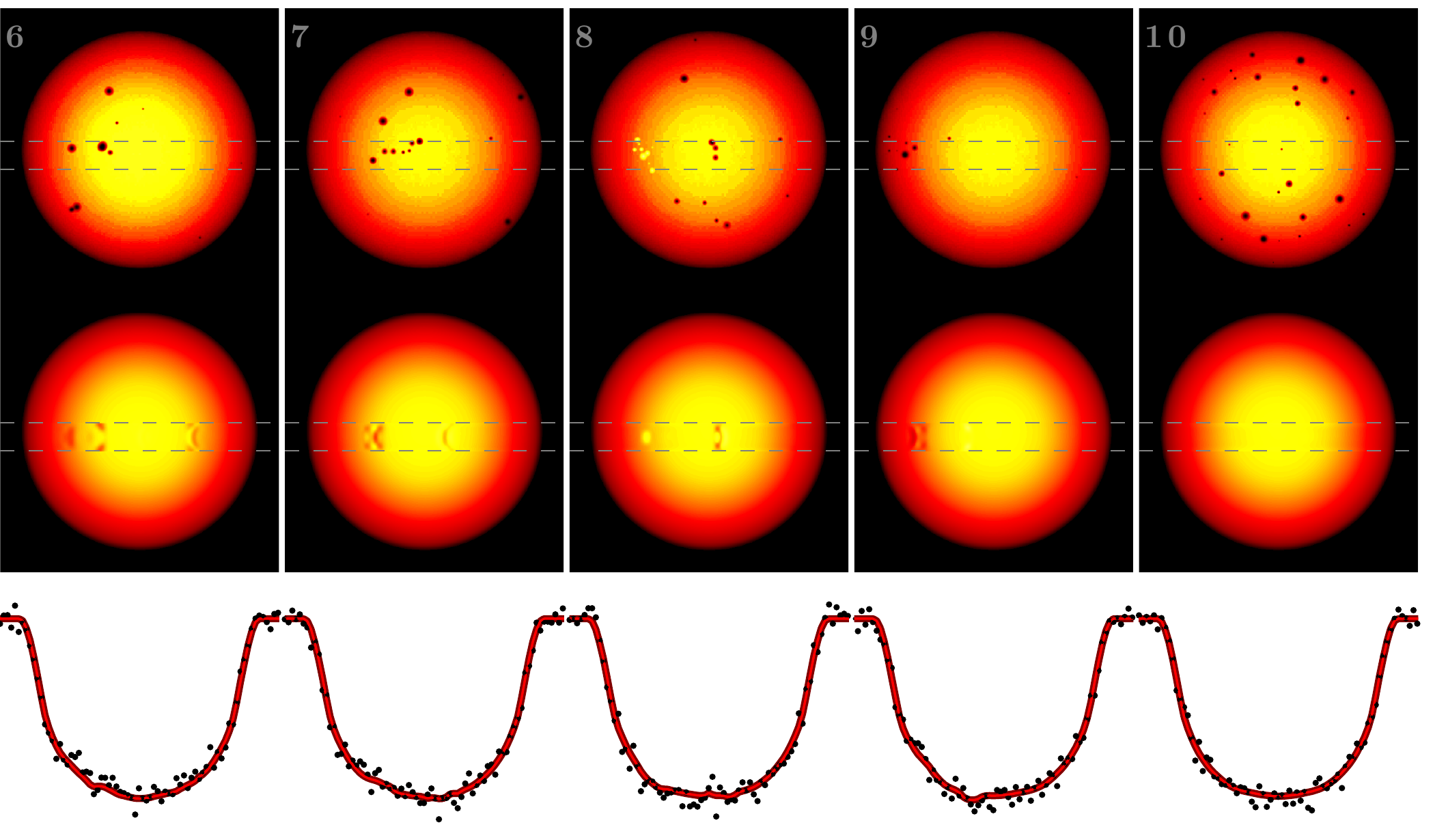}
        \sidecaption
        \caption{Ten simulated transit observations were fed through the transit imaging method described in this paper. For each transit event we show three rows. \newline
        First row from the top: Original stellar surface map that was used to simulate an observation.  \newline
        Second row: Reconstructed stellar surface map. \newline
        Third Row: Light curves. The "observations" are shown as black dots, and the synthetic light curves crated from the reconstructed surface map are shown as solid red lines.\newline
        In these tests the method successfully recovered all strong brightness variations that were crossed by the planet during a transit. The weak spots and the spots outside the transit chord were not recovered. }
        \label{results1}
\end{figure*}

We tested the method by applying it to ten simulated observations of the same star, see the bottom panels of Fig.~\ref{results1}. The S/N and exposure times of the simulated data were similar to what is expected from a typical light curve from the Transiting Exoplanet Survey Satellite (TESS) \citep{Ricker2015JATIS}. We first established the specific intensity of the spot-free star by phase-folding and median-combining the ten light curves as described in Sec.~\ref{SpotFreeSpecificIntensity}. The surface map for each individual transit was then reconstructed using the iterative method described in Sec.~\ref{MappingStellarSurface}. The regularization parameter $\Lambda$ was chosen so that the false-detection rate is 5\%, as described in see Sec.~\ref{RegularizationMethod}. The recovered original maps are shown in Fig.~\ref{results1}. All larger spots or groups of spots that the planet crosses during transit were recovered in the test.

\subsection{Test of the method on individual transits} \label{Test of method on individual transit}
Here the transit imaging method is applied to transit observations made with the FOcal Reducer and low dispersion Spectrograph 2 (FORS2) \citep{Appenzeller1998Msngr} at the VLT. Observations were taken from the ESO archive. We used observations of five systems: GJ 1214, GJ 436, WASP-17, WASP-43, and WASP-80. We reduced the observations in the same way as in \citet{Aronson2018AJ}. To achieve a sufficient S/N, we sorted all the wavelengths into three equal size bins: blue, green, and red. The intensities in each bin were averaged and the time sequences were analyzed separately as if we had three transit light curves. In addition to the color light curves, we also averaged over the whole wavelength range to create a single light curve with a high S/N. The maps created from the red, green, and blue light curves also allowed for the creation of colorized stellar surface maps by recombining them as a separate color-channels to form the final map. 

For some systems, data of sufficiently high quality were only available for a single transit event. This was the case for GJ 436, WASP-17, and WASP-43. For these the reference specific intensity could not be established from medium-combining phase-folded light curves. Instead, we created this reference-specific intensity by removing data with large deviations form a typical transit-light-curve shape. Removed data were primarily exposures with poor quality and suspected spot crossing events. The specific intensity function was then fitted to this. For GJ 1214 and WASP-80, observations of multiple transit events were available, and we were able to apply the method as described in Sec.~\ref{SpotFreeSpecificIntensity}.

The orbital parameters used during the phase-folding and determination of the planet position during each exposure were taken from the Exoplanet Orbit Databaset \citep{2014PASP..126..827H}. The stellar parameters shown in Table.~\ref{stellarParameters} were taken from the same place, although we note that they were not used to map the stellar surface.

\begin{table}[] \label{stellarpara}
        \centering
        \caption{Stellar parameters of the stars observed by FORS2.}
        \label{stellarParameters}
        \begin{tabular}{lllll}
                \hline \hline
                & Radius & $\mathrm{T_\mathrm{eff}}$ & V mag & log(g)\\
                & ($\mathrm{R_{\odot}}$) & (K)    &     & cgs\\
                \hline
                GJ 1214 & 0.21 & 3030 & 15.1  & 4.991\\
                GJ 436  & 0.46 & 3350 & 10.7  & 4.843\\
                WASP-17 & 1.20 & 6550 & 11.6  & 4.20\\
                WASP-43 & 0.60 & 4400 & 12.4  & 4.646\\
                WASP-80 & 0.57 & 4150 & 11.7  & 4.60\\
                \hline
        \end{tabular}
\end{table}

\subsubsection{Results - GJ 1214} \label{GJ1214_d}
\begin{figure*}
        \centering
        \includegraphics[width=\hsize]{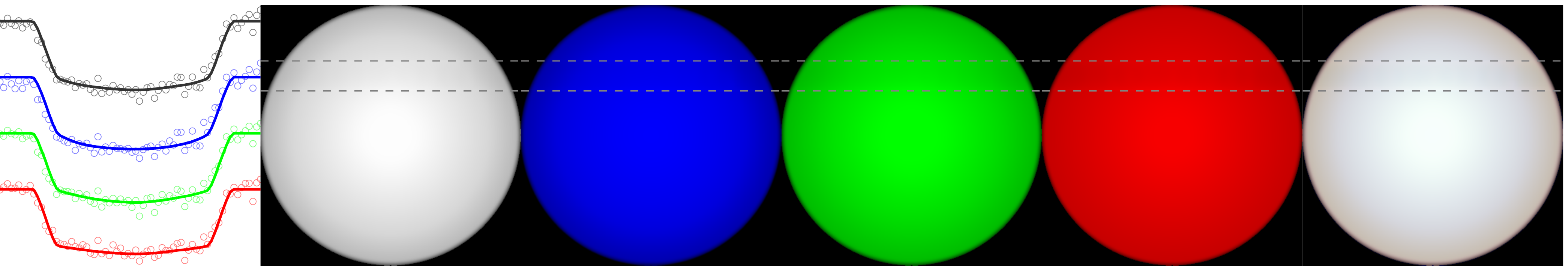}
        \includegraphics[width=\hsize]{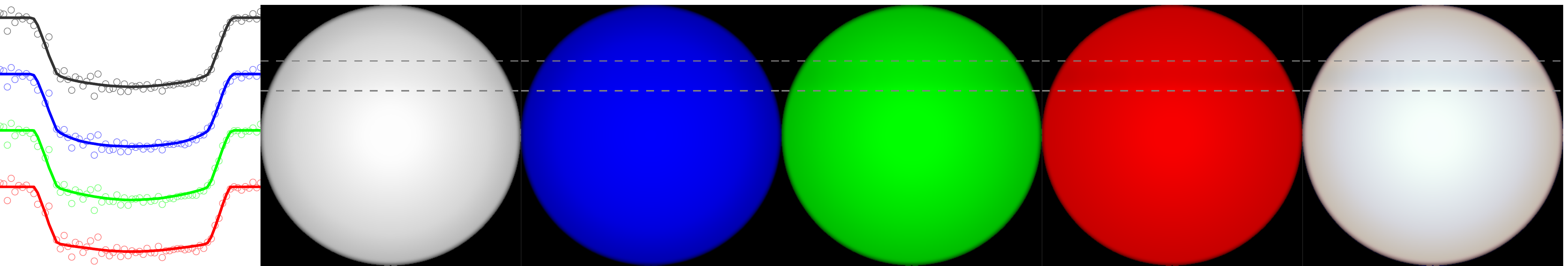}
        \includegraphics[width=\hsize]{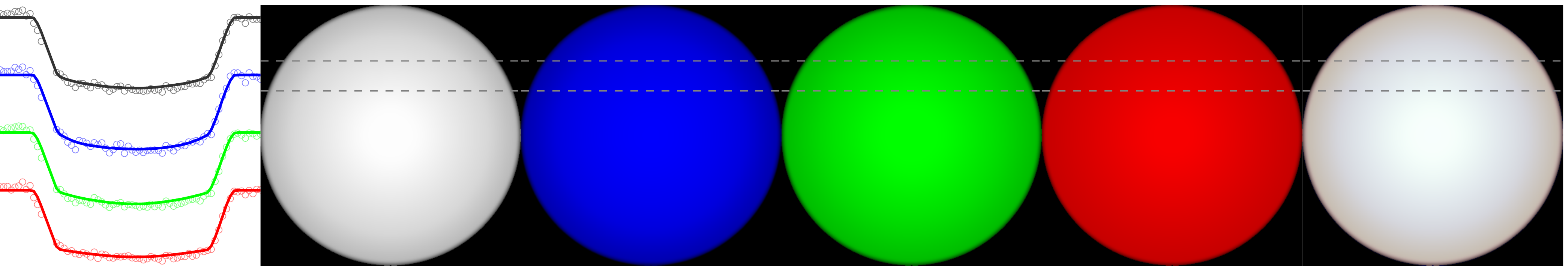}
        \includegraphics[width=\hsize]{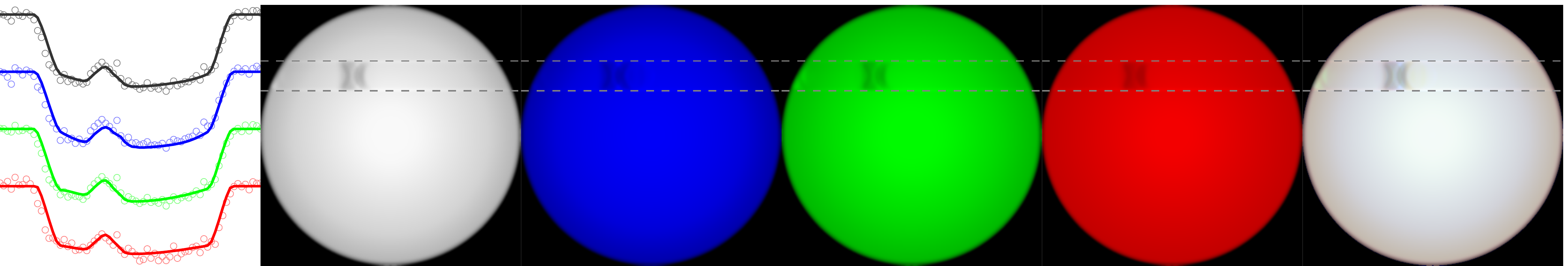}
        \includegraphics[width=\hsize]{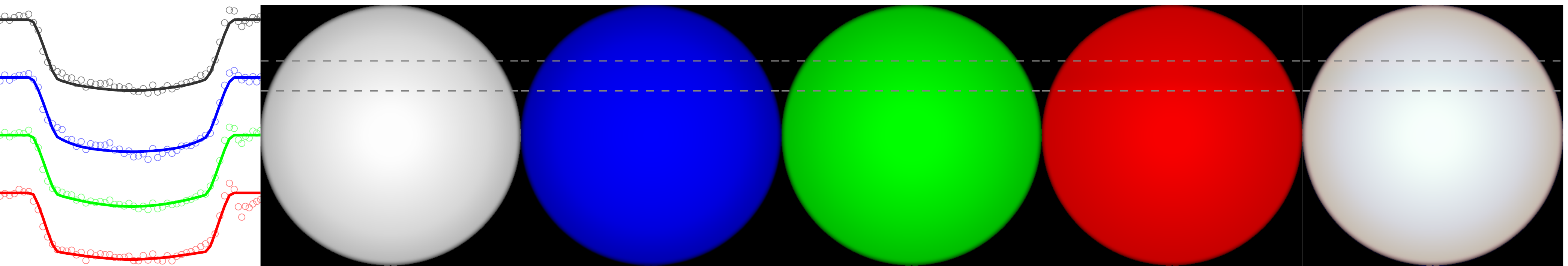}
        \includegraphics[width=\hsize]{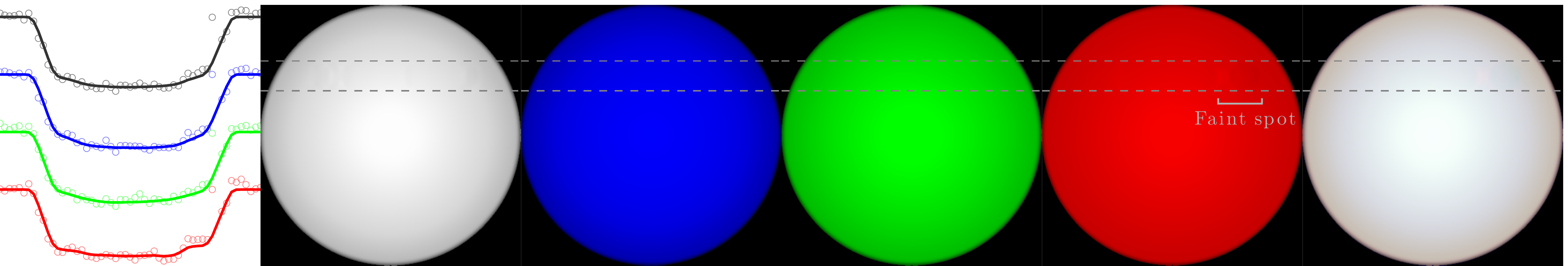}
        \includegraphics[width=\hsize]{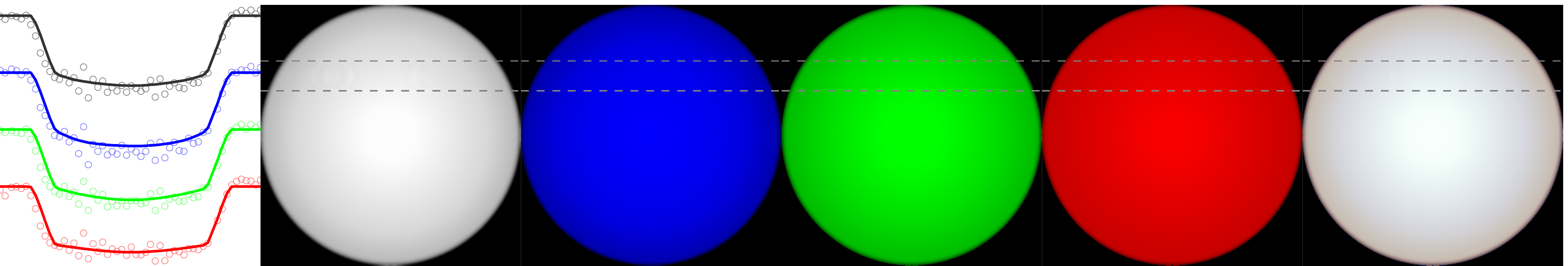}
\caption{Recovered maps of GJ 1214. Each row is a single transit event that is divided into five panels. \newline
        The first panel from the left shows the transit light curves. Circles show observed data, and the solid lines present the light curve produced by the recovered maps. There are four light curves in each plot. The upper curve in gray shows observations from all wavelengths, while the lower three curve show the data from the three wavelength bins (short wavelengths in blue, intermediate in green, and long in red). \newline
        The second to fifth panels show the reconstructed surface maps in the same wavelength bins as the light curves. From left to right: all wavelengths, short wavelengths, intermediate wavelengths, and long wavelengths. The rightmost stellar map is a color-composite of the blue, green, and red maps. In all maps the dashed gray lines show the transit chord. Observations of this system were made in 0.73-1.02 $\mu m$. A dark spot (or group of spots) was detected in the fourth transit. A faint spot is detected in the red wavelengths of the sixth transit.}
        \label{GJ1214_combo}
\end{figure*}

GJ 1214 is a small and cool star (the radius is $0.21 R_\odot$ and $T_\mathrm{eff} = 3030$) with a transiting super-Earth \citep{2009Natur.462..891C}. It is an extensively studied system because its planetary-to-stellar radius ratio is favorable. Only a few other systems with transiting super-Earths have such a high ratio, and this makes this system one of the best for studying super-Earths. Spots on this star have previously been detected in transit light curves \citep{2011ApJ...736...12B, 0004-637X-730-2-82}. For this test, observations from seven transits were used, in which evidence of one large spot during the forth transit was recovered. A weak spot can also barely be seen in the sixth transit. The latter is only detected in the red wavelength bin where the S/N is lowest. Because spots are expected to have high contrast at shorter wavelengths, this is probably an artifact of observations and not a real feature. Our maps support the opinion that the observed stellar brightness variations of up to 1\% can be attributed to variation of spot coverage. The resulting maps are presented in Fig.~\ref{GJ1214_combo}. 

\subsubsection{Results - GJ 436} \label{GJ436_d}
\begin{figure*}
        \centering
        \includegraphics[width=\hsize]{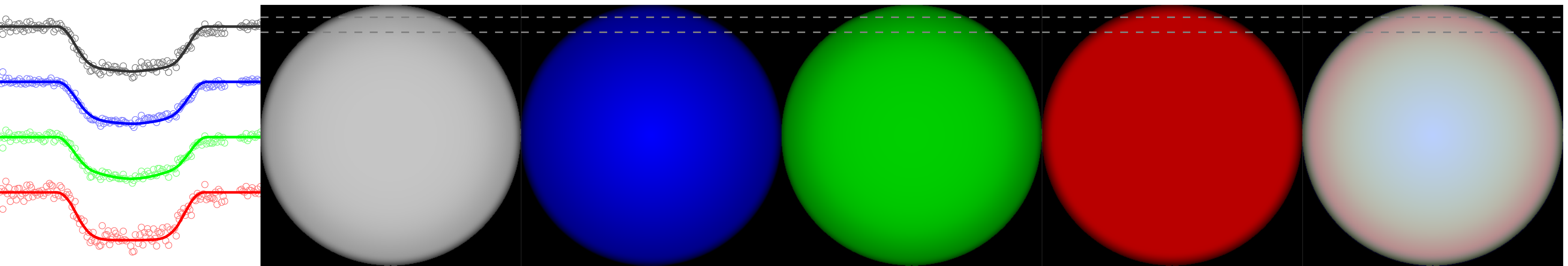}
        \caption{Recovered maps of GJ 436. See Fig.~\ref{GJ1214_combo} for a description of the figure. \newline
                Observations of this system were made in 0.78-0.95 $\mu m$.  \newline
                No spots or other brightness irregularities were detected on this star.}
        \label{GJ436_combo}
\end{figure*}

GJ 436 has a radius of $0.46 R_\odot$ and a hot Neptune \citep{2004ApJ...617..580B} that transits it. The planet transits the star very close to the limb, which makes the area of the star that can be mapped relatively small. No significant departures from a spot-free stellar surface were detected. Results are shown in Fig.~\ref{GJ436_combo}. 

\subsubsection{Results - WASP-17} \label{WASP-17_d}
\begin{figure*}
        \centering
        \includegraphics[width=\hsize]{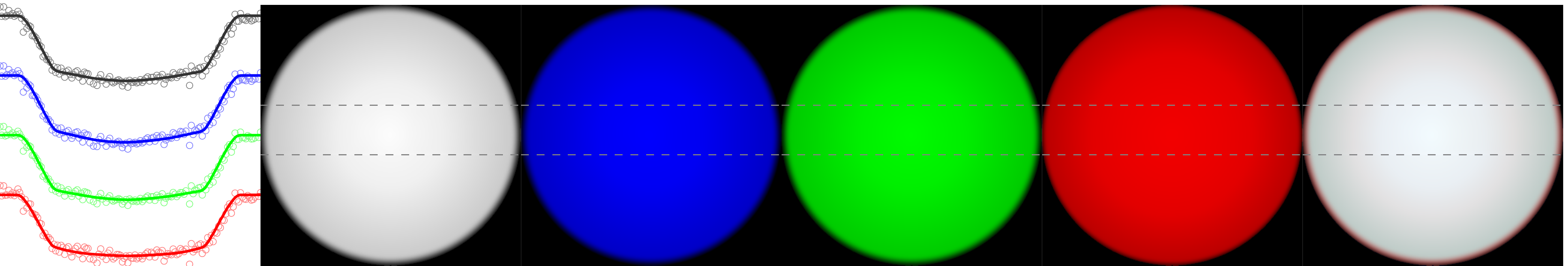}
        \caption{Recovered maps of WASP-17. See Fig.~\ref{GJ1214_combo} for a description of the figure. \newline
                Observations of this system were made in 0.74-1.06 $\mu m$.  \newline
                No spots or other brightness irregularities were detected on this star. }
        \label{WASP-17_combo}
\end{figure*}

The large size of WASP-17b relative to its host star (the stellar radius is $1.20 R_\odot$ and the planet radius is $1.93 R_{Jupiter}$ \citep{2010ApJ...709..159A}) combined with an orbital projection that crosses the center of the stellar disk enables us to probe a large portion of the stellar surface (24\%). No brightness variations in the transited region of the star were detected in these data, however. The resulting maps and light curves are shown in Fig.~\ref{WASP-17_combo}. 

\subsubsection{Results - WASP-43} \label{WASP-43_d}
\begin{figure*}
        \centering
        \includegraphics[width=\hsize]{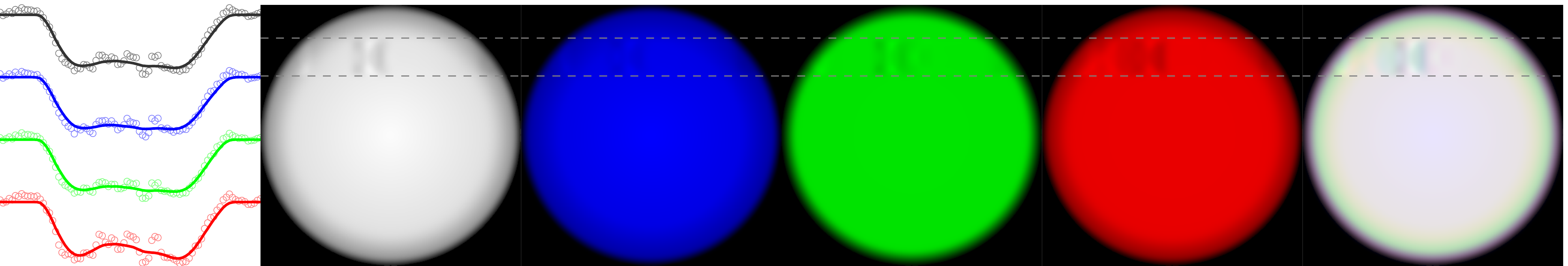}
        \caption{Recovered maps of WASP-43. See Fig.~\ref{GJ1214_combo} for a description of the figure. \newline
                Observations of this system were made in 0.56-0.85 $\mu m$. \newline
                A band of spots was detected in the central parts of the stellar disk. }
        \label{WASP-43_combo}
\end{figure*}

WASP-43 is a small star with a radius of $0.6 R_\odot$. It has a transiting hot Jupiter \citep{2011A&A...535L...7H}. The transit light curve shows clear signs of brightness irregularities on the stellar surface in the central region of the light curve. Our method interprets this as a group of dark spots. The results are shown in Fig.~\ref{WASP-43_combo}. 

\subsubsection{Results - WASP-80} \label{WASP-80_d}
\begin{figure*}[h]
        \centering
        \includegraphics[width=\hsize]{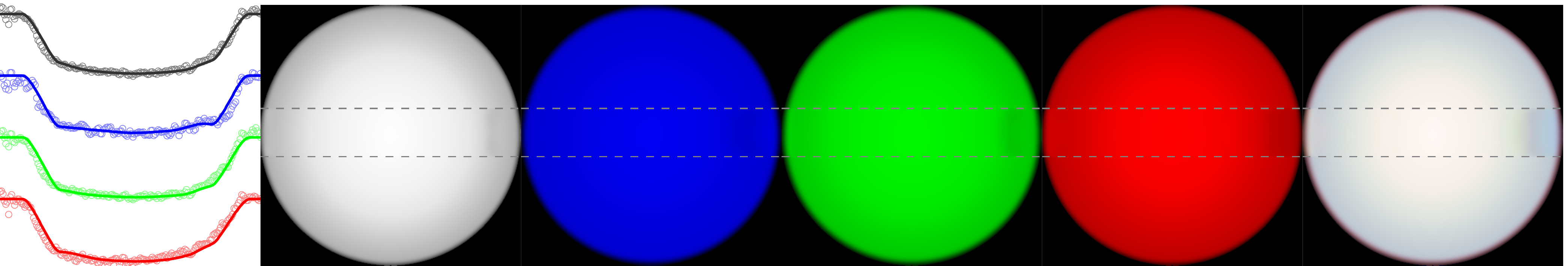}
        \includegraphics[width=\hsize]{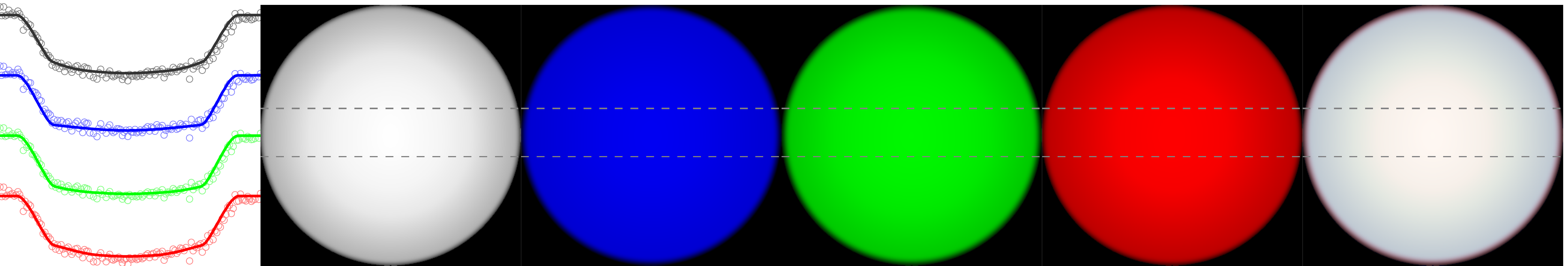}
        \caption{Recovered maps of WASP-80. See Fig.~\ref{GJ1214_combo} for a description of the figure. \newline
Observations of this system were made in 0.74-1.06 $\mu m$. \newline
In the recovered maps of the first transit, we detect a large dark spot close to the limb of the star.}
        \label{WASP-80_combo}
\end{figure*}

Similarly to WASP-17, the size ratio of WASP-80 to its transiting hot Jupiter \citep{2013A&A...551A..80T} allows for mapping of a large portion of the stellar disk (23\%), including the most central regions. However, the star and the planet are only half the size of their counterparts in WASP-17 (the stellar radius is $0.57 R_\odot$ and the planet radius is $0.95 R_{Jupiter}$). For this system we had access to observations from two transit events, and in one of them, a large but faint dark spot close to the limb of the star was detected. This might be a group of smaller spots rather than a single large spot. As seen in Fig.~\ref{results1}, transit 5, a tight group of small spots can be interpreted as a single larger spot. The resulting maps are presented in Fig.~\ref{WASP-80_combo}.

\subsection{Method comparison - PRISM} \label{prism_comp}
In this section we compare our method for mapping stellar surfaces with the method presented by \citet{Tregloan-Reed2013MNRAS}, the PRISM code. Their method is applied to individual transit light curves and requires no input from previous studies. It fits orbital parameters, planetary size, and spots all at once, thus eliminating potential degeneracy between the presence of spots and other parameters. Four free parameters are connected to the spots in the PRISM code: spot position (longitude and latitude of its center), spot radius, and spot brightness contrast. This means that spots are assumed to be circular and of uniform brightness.

\begin{figure}[h]
        \includegraphics[width=\hsize]{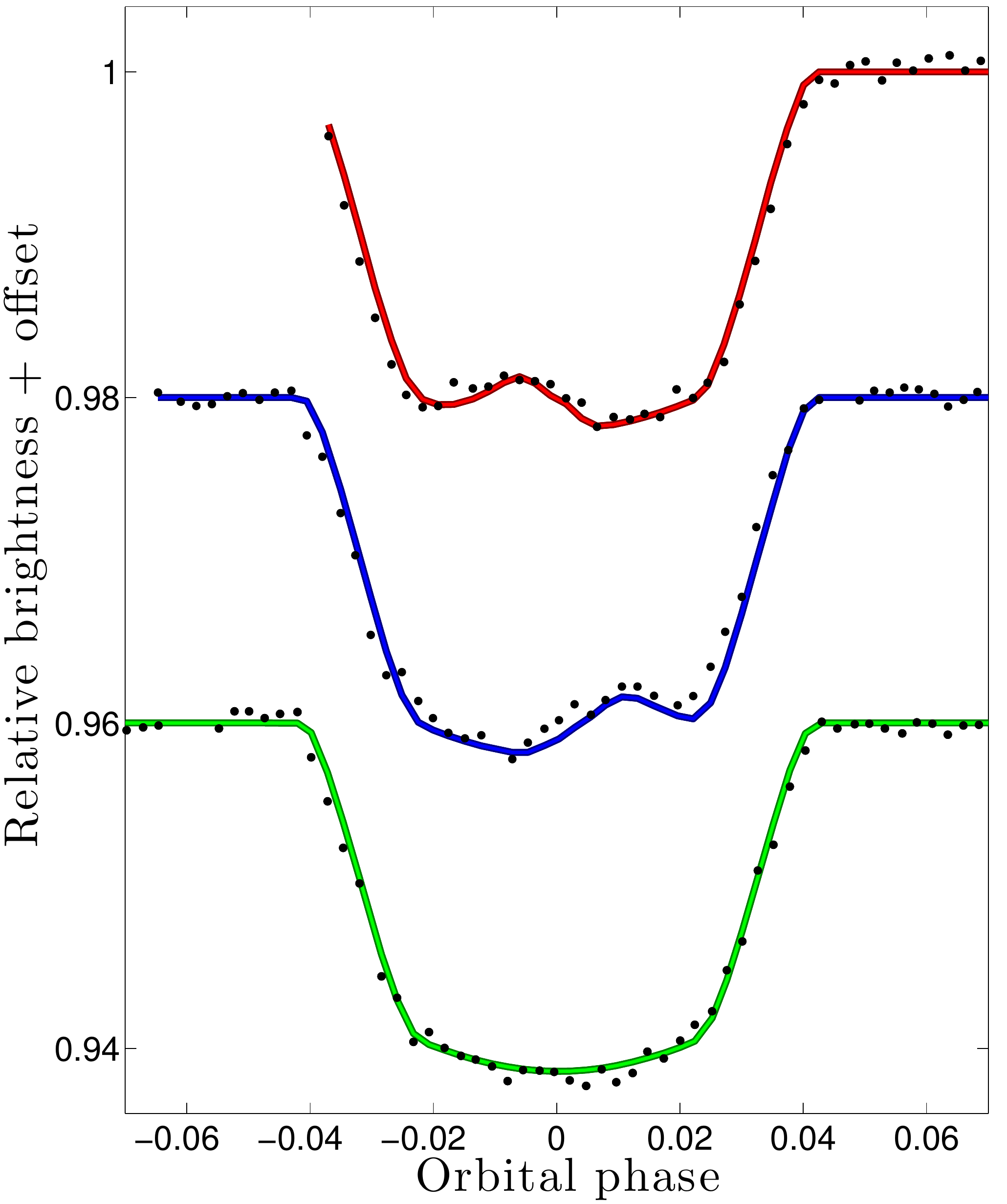}
        \includegraphics[width=\hsize]{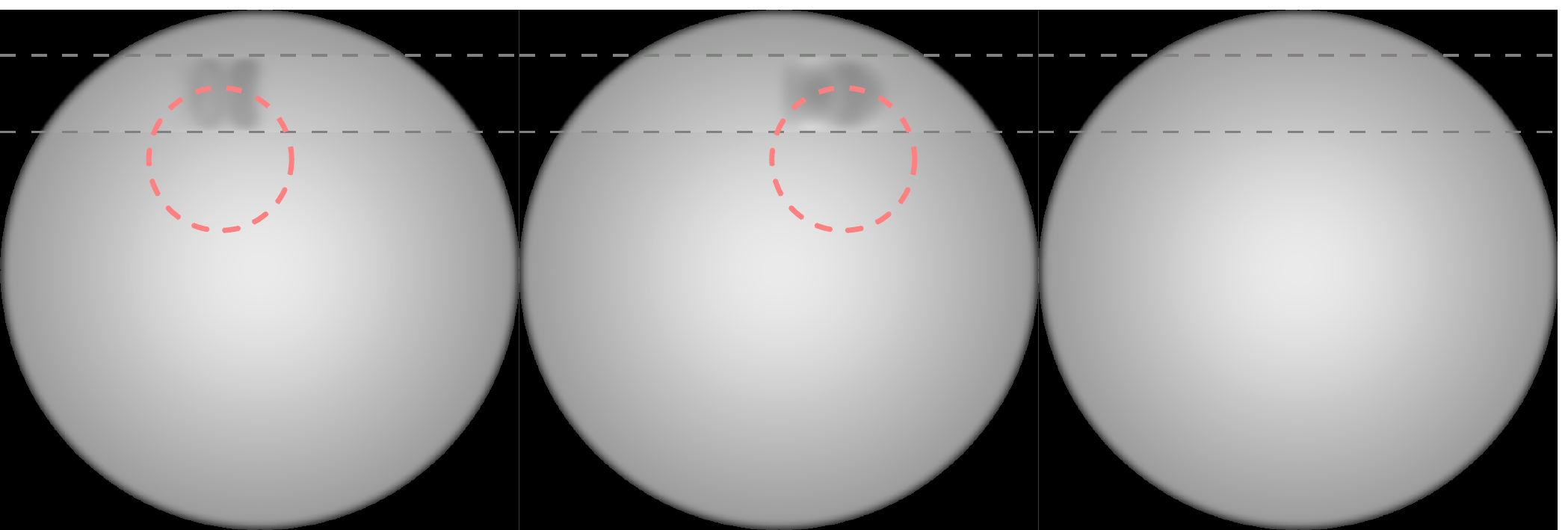}
        \caption{Our method applied to WASP-19 data. This is the same data as were used to test the PRISM code \citep{Tregloan-Reed2013MNRAS}. \newline
                        Upper panel: Observed light curves of three transit events shown as black dots, and the light curves produced by our recovered stellar maps is shown as solid lines. \newline
                        Bottom panels: Recovered stellar maps. The left map corresponds to the red light curve in the upper panel. The central map corresponds to blue light curve, and the right map corresponds to the green light curve. Star spots were detected in the recovered maps from the first two transit events. The dashed red circle shows the size and position of the spots detected with the PRISM code.}
        \label{wasp19comp}
\end{figure}

We compared our method with the PRISM code by analyzing the exact same data set that was previously used for testing the PRISM code: three transits of WASP-19 b. See \citet{Tregloan-Reed2013MNRAS} for details. We took the orbital parameters and planet-to-star radius ratio from the same paper. However, our method does not require any input on stellar parameters. After analysis, our method detects spots in the same light curves and at the same longitudes, see Fig.~\ref{wasp19comp}. Comparing latitudes of the apparent spot centers is non-trivial because PRISM makes assumptions on the spot outside the transit chord, while our method only attempts to fit the part of the spot that is located on the transit chord. As a result, PRISM finds that the detected spots are larger than the width of the transit chord, and only a fraction of the spots are crossed by the transiting planet. Our recovered maps yield no information about part of the spots outside the transit chord. When only the part of the spots on the transit chord is considered, the maps from the PRISM code and what we recover is consistent. The spots we recover are not circular and potentially consist of several smaller spots close to each other.

\section{Conclusions} \label{Conclusions}
The transit imaging technique presented in this paper allows reconstructing brightness variations on stellar surfaces, with no dependencies on either stellar atmosphere models or limb-darkening equations. When observations include multiple transit light curves of the same system, the analysis can be fully automated and requires no additional inputs beyond the light curves themselves and orbital parameters, thus allowing for consistent application on all observed transit light curves. This makes the method well suited for survey-type observations because such surveys typically include many transit light curves of the same system, but also because the number of transiting systems that is expected to be available will enable the creation of a database of stellar spot coverage that is large enough to allow a statistical analysis of the field. This could give the scientific community the tools to answer questions such as how many stars have spots, and whether the time base of observations is long enough to detect systematic trends in the spot coverage. If the orbital inclination of the planaet is known with respect to the stellar equator, it might be possible to determine at what latitudes spots typically appear. Other possibilities might be to find correlations between stellar parameters such as age, type, or metallicity with the likelihood of the star having spots, their contrasts, or their lifetimes.

The regularization technique that is used in this paper is also well suited for creating a database of stellar spot coverage. By tuning the regularization parameter, the method sensitivity and frequency of false positives can be controlled. If the end user requires a high sensitivity, or alternatively a low false-detection rate, this can easily be achieved. The rate of false detections is estimated by applying the method to synthetic data of spotless stars. The fraction of stellar maps recovered from this set of simulations gives the rate of false detection. If a star has a variable flux, one can search for connections between spot coverage and variability in stellar flux. 

Access to transit survey data also allows for some unique opportunities that high-precision observations of a single transit event do not allow. For example, when a spot is discovered and the planet orbits in the same plane as the stellar rotation and the orbital period is shorter than both the rotational period of the star and the lifetime of the spot, it is possible to follow the spot movement as the star rotates. This can allow estimating the stellar rotational period, its plane of rotation, and the lifetime of the spot. If the star has several transiting planets and they transfer at different impact parameters, the region of the stellar disk that can be reconstructed is increased. 

The method was tested on a set of synthetic data of survey type, meaning several observed transit events at a stable S/N. The results are promising. All strong spots or groups of spots were successfully recovered in the reconstructed maps of the stellar disks. The method was also tested on a set of observational data from the ground-based medium-resolution spectrograph FORS2 at the VLT. In this case, the number of available transit observations was  smaller than what is needed to create a meaningful database. Observations were also made over the course of several years, which precludes tracking the spot movement. The method was still able to provide maps of the stellar disks as snapshots. Spots were detected in three of the five observed stars. 

Data from the TESS \citep{Ricker2015JATIS} are expected to provide the required quality and volume to enable the creation of a database of stellar spot coverage. The survey space telescope PLAnetary Transits and Oscillations of stars (PLATO) \citep{2016SPIE.9904E..28R} will later allow adding more stars to this database.

\begin{acknowledgements}
        This research has made use of the Exoplanet Orbit Database and the Exoplanet Data Explorer at exoplanets.org.\\
        The authors would also like to acknowledge the ESO archive for providing a complete data set for\\
        VLT FORS2 observations of\\
        - GJ 1214 (284.C-5042(B), 285.C-5019(A), 285.C-5019(C), 087.C-0505(A), 089.C-0020(I), 089.C-0020(D), 089.C-0020(B)), \\
        - GJ 436 (091.C-0222(A)),\\
    - WASP-17 (087.C-0225(C)), \\
    - WASP-43 (094.C-0639(B)), and \\
    - WASP-80 (091.C-0377(C), 091.C-0377(E)).\\
    EFOSC observations of\\
    - WASP-19: 084.D-0056(A).
\end{acknowledgements}

\bibliography{mybib12.bib}
\bibliographystyle{aa}

\end{document}